\documentclass[11pt,a4paper,superscriptaddress,aps,preprint]{revtex4}
\usepackage{amsmath}
\usepackage{amssymb}
\usepackage{graphicx}
\makeatletter
\newcommand{\bea}{\begin{eqnarray}}
\newcommand{\eea}{\end{eqnarray}}

\newcommand{\pa}{\partial}

\begin{document}

\title{On the Horava-Lifshitz-like Gross-Neveu model}

\author{A. M. Lima}
\affiliation{Departamento de F\'{\i}sica, Universidade Federal da Para\'{\i}ba, Caixa Postal 5008, 58051-970, Jo\~ao Pessoa, Para\'{\i}ba, Brazil}
\email{amlima,jroberto, petrov, rfreire@fisica.ufpb.br}

\author{T. Mariz}
\affiliation{Instituto de F\'\i sica, Universidade Federal de Alagoas, 57072-270, Macei\'o, Alagoas, Brazil}
\email{tmariz@fis.ufal.br}

\author{R. Martinez}
\affiliation{Departamento de Ciencias F\'{i}sicas, Facultad de Ingenier\'{i}a y Ciencias, Universidad de la Frontera, Casilla 54-D, Temuco, Chile}
\email{ricardo.martinez@ufrontera.cl}

\author{J. R. Nascimento}
\affiliation{Departamento de F\'{\i}sica, Universidade Federal da Para\'{\i}ba, Caixa Postal 5008, 58051-970, Jo\~ao Pessoa, Para\'{\i}ba, Brazil}
\email{amlima,jroberto, petrov, rfreire@fisica.ufpb.br}

\author{A. Yu. Petrov}
\affiliation{Departamento de F\'{\i}sica, Universidade Federal da Para\'{\i}ba, Caixa Postal 5008, 58051-970, Jo\~ao Pessoa, Para\'{\i}ba, Brazil}
\email{amlima,jroberto, petrov, rfreire@fisica.ufpb.br}

\author{R. F. Ribeiro}
\affiliation{Departamento de F\'{\i}sica, Universidade Federal da Para\'{\i}ba, Caixa Postal 5008, 58051-970, Jo\~ao Pessoa, Para\'{\i}ba, Brazil}
\email{amlima,jroberto, petrov, rfreire@fisica.ufpb.br}

\begin{abstract}
We describe a Horava-Lifshitz-like reformulated four-fermion Gross-Neveu model describing the dynamics of two-component spinors in $(2+1)$-dimensional space-time. Within our study, we introduce the Lagrange multiplier, study the gap equation (including the finite temperature case) which turns out to display essentially distinct behaviors for even and odd values of the critical exponent $z$, and show that the dynamical parity breaking occurs only for the odd $z$.  {We demonstrate that for any odd $z$, there exists a critical temperature at which the dynamical parity breaking disappears. Besides of this, we} obtain the effective propagator and show that the resulting effective theory is renormalizable within the framework of the $\frac{1}{N}$ expansion for all values of $z$. As one more application of the dynamical parity breaking, we consider coupling of the  vector field to the fermions in the case of a simplified spinor-vector coupling and discuss the generation the Chern-Simons term.
\end{abstract}

\maketitle

\section{Introduction}

The Horava-Lifshitz (HL) approach \cite{Hor} has recently acquired a great scientific attention. This approach is characterized by the essential asymmetry between space and time coordinates: the equations of motion of the theory
are invariant under the rescaling $x^i\to bx^i$, $t\to b^zt$, where $z$,
the critical exponent, is a number characterizing the
ultraviolet behavior of the theory. The main reason for it is that the HL-like reformulation of the known field theory models with a nontrivial critical exponent $z>1$ an essential improvement of the renormalization behavior of these models takes place. In particular, the four-dimensional gravity becomes renormalizable at $z=3$. Different issues related to the HL gravity, including its cosmological aspects \cite{HorCos}, exact solutions \cite{Lu}, black holes \cite{BH} 
were considered in a number of papers. 

At the same time, study of the impacts of the HL extension of other field theory model is a very interesting problem. Some aspects of the HL generalizations for the gauge field theories were presented in \cite{ed}. Renormalizability of the HL-like scalar field theory models has been discussed in details in \cite{Anselmi}, and explicit study of their renormalization was carried out in \cite{PRMG}.  The Casimir effect for the HL-like scalar field theory has been considered in \cite{ourcas}. In \cite{cpn}, the HL modifications of the $CP^{N-1}$ theory were studied. The purely scalar analogue of the sigma-model has been discussed in \cite{prsmg}. Different examples of calculating the effective potential in HL-like theories involving scalars coupled to gauge and/or spinor fields are presented in \cite{ourEP}.

Many other interesting aspects of HL-like fermionic theories of spinor fields can be studied as well. Indeed, as it is well known, namely different couplings within the four-fermion model give origin to the phenomenon of the emergent dynamics which possibly allows to treat the photon as a Goldstone boson and to develop other phenomenologically interesting concepts \cite{Bjorken}. Actually, the original Gross-Neveu model \cite{GN} also was introduced in order to study the emergent dynamics, and, to be more precise, to discuss the dynamical parity breaking taking place due to arising of the nontrivial mass term for spinors. Therefore, it would be natural to generalize studies of emergent dynamics and dynamical parity breaking for the HL-like theories with a nontrivial $z$. Some preliminary studies of this issue have been carried in \cite{Dhar}. 

Here, we will study the contributions of HL-like spinor fields to an effective dynamics for the  Lagrange multiplier, thus establishing the HL-like Gross-Neveu model, and discuss its renormalizability within the framework of the  $\frac{1}{N}$ expansion. As one more illustration, we consider the case when the HL-like spinor field is coupled to the vector field. For this coupling, the Maxwell-like terms were earlier generated in \cite{ourQED}, and it remains to study the possibility to carry out the perturbative generation of the Chern-Simons term as well, which we perform in this paper for a some simplified theory. 

The structure of the paper looks like follows. In the section 2 we present the HL-like Gross-Neveu model. In the section 3 we obtain the effective propagators and gap equation or the Lagrange multiplier and prove the renormalizability of the resulting theory. In the section 4, we carry out the perturbative generation of the Chern-Simons term. Finally, in Summary we discuss our results and perspectives.

\section{Gross-Neveu model}

In this paper, the main object of our study is the Gross-Neveu model \cite{GN,sigma}. Its natural HL-like generalization, for the set of $N$ spinor fields, looks like
\bea
\label{acspi1}
S=\int dt d^2x
\sum\limits_{j=1}^N\left[\bar{\psi}_j(i\gamma^0\pa_0+(i\gamma^i\pa_i)^z)\psi_j-\frac{g}{2N}(\bar{\psi}_j\psi_j)^2\right],
\eea
where $j=1,2,\cdots,N$.  
The number of spatial derivatives acting on the spinor fields is equal to $z$,
that is, the critical exponent for the spinor fields. We will study the effective dynamics of this model within the framework of the $\frac{1}{N}$ expansion, following the lines developed in \cite{GNR} for the Lorentz invariant case,  and use conventions adopted in \cite{GNR}.

First of all,  we introduce the Lagrange multiplier, that is, an auxiliary scalar field $\sigma$ allowing to avoid the four-fermion coupling, so that the action takes the form
\bea
S=\int dt d^2x
\left[\sum\limits_{j=1}^N\left[\bar{\psi}_j\left(i\gamma^0\pa_0+(i\gamma^i\pa_i)^z-\sigma\right)\psi_j\right]+\frac{N}{2g}\sigma^2\right].
\eea
Now, we introduce a mass through shift: $\sigma\to\sigma+m^z$ with $m$ has the mass dimension 1, so that the vacuum expectation of $\sigma$ turns out to be equal to zero. We get
\bea
\label{act}
S=\int dt d^2x
\left[\sum\limits_{j=1}^N\left[\bar{\psi}_j(i\gamma^0\pa_0+(i\gamma^i\pa_i)^z-m^z-\sigma)\psi_j\right]+\frac{N}{2g}(\sigma^2+2\sigma m^z)\right].
\eea
Here we disregarded the constant, field-independent additive term.
The mass term $m^z \bar{\psi}_j\psi_j$ breaks the parity \cite{GNR}. Indeed, under the parity transformation, that is, inversion of the axis $x_1$, of the form
\bea
\psi(x_0,x_1,x_2)\to\gamma^1\psi(x_0,-x_1,x_2)
\eea
the mass term changes the sign since $(\gamma_1)^2=-{\mathbf 1}$ (cf. \cite{GNR}); we remind that $\bar{\psi}=\psi^{\dagger}\gamma^0$. One can note as well that under the parity transformation, the time part of the kinetic term $\bar{\psi}\gamma^0\pa_0\psi$ is invariant, and its spatial part $\bar{\psi}(\gamma^i\pa_i)^z\psi$ is invariant for the odd $z$, including the Lorentz-invariant case $z=1$, and changes the sign for the even $z$; however,  namely the case of the odd $z$, where the kinetic term is completely parity invariant, and the mass term breaks the parity, is the most interesting one for us since just in this case the dynamical mass generation occurs, as we show further.

The propagator of the $\psi$ field is
\bea
\label{3}
S_{ij}(k)=<\bar{\psi}_i(k)\psi_j(-k)>=i\frac{\delta_{ij}}{\gamma^0k_0+(\gamma^ik_i)^z-m^z}\equiv\delta_{ij}S(k).
\eea
We use the signature $(+--)$. Now, one should emphasize an important difference between odd and even $z$. Indeed, for even $z=2n$, one has $(\gamma^ik_i)^{2n}=(-\vec{k}^2)^n$, and for odd $z=2n+1$, one has $(\gamma^ik_i)^{2n+1}=\gamma^ik_i(-\vec{k}^2)^n$. Then, we have
\bea
z&=&2n: \quad\,  S(k)=i\frac{\gamma^0k_0-[(-\vec{k}^2)^n-m^z]}{k^2_0-[(-\vec{k}^2)^n-m^z]^2}\equiv i\frac{\gamma^0k_0-\omega(\vec{k})}{k^2_0-\omega^2(\vec{k})},\nonumber\\
z&=&2n+1: \quad\, S(k)=i\frac{\gamma^0k_0+\gamma^ik_i(-\vec{k}^2)^n+m^z}{k^2_0-(\vec{k}^2)^z-m^{2z}}\equiv i\frac{\gamma^0k_0+\gamma^ik_i(-\vec{k}^2)^n+m^{2n+1}}{k^2_0-\Omega^2(\vec{k})}.
\eea
These propagators will be further used for studying of quantum corrections in our theory.

\section{Dynamics of the Lagrange multiplier}

\subsection{Gap equation in the zero temperature case}

To study the dynamics of the Lagrange multiplier $\sigma$, we start with the gap equation, i.e. let us impose a condition that the vacuum expectation $<\sigma>=0$. Graphically, it is represented by the Feynman diagram given at Fig. 1.

\begin{figure}[htbp] 
\includegraphics[scale=0.8]{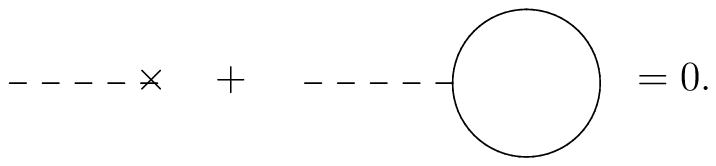}
\caption{\small{Gap equation.}}
\label{figure1}
\end{figure}

\noindent Here the sign $\times$ means the constant $m^z$ originated from the action (\ref{act}), the solid line is for $\psi$ field, and the dashed one is for $\sigma$ field. This equation means that, given the effective potential 
\bea
V_{eff} = -\frac{\sigma^2}{2g}+i\mathrm{tr} \int \frac{dk_0d^2\vec{k}}{(2\pi)^3}
\ln\left(\gamma^0k_0+(\gamma^ik_i)^z-\sigma\right),
\eea
we are interested in finding the minima, that is, in solutions of the expression 
\bea
\frac{dV_{eff}}{d\sigma}\Big|_{\sigma=m^z}=-\frac{m^z}{g}-i\mathrm{tr} \int \frac{dk_0d^2\vec{k}}{(2\pi)^3} \frac{1}{\gamma^0k_0+(\gamma^ik_i)^z-m^z} = 0.
\eea
Then, after Wick rotation, $k_0 \to ik_0$, for even $z=2n$, we have
\bea
\frac{m^z}{g}=-{\rm tr}\int\frac{dk_0d^2\vec{k}}{(2\pi)^3}\frac{i\gamma^0k_0-[(-\vec{k}^2)^n+m^z]}{k^2_0+[(-\vec{k}^2)^n+m^z]^2},
\eea
and for odd $z=2n+1$,
\bea
\label{12}
\frac{m^z}{g}=-{\rm tr}\int\frac{dk_0d^2\vec{k}}{(2\pi)^3}\frac{i\gamma^0k_0+\gamma^ik_i(-\vec{k}^2)^n+m^z}{k^2_0+(\vec{k}^2)^z+m^{2z}}.
\eea
Calculation of the trace is straightforward, so that for $z=2n$, we get
\bea
\frac{m^z}{g}=2\int\frac{dk_0d^2\vec{k}}{(2\pi)^3}\frac{[(-\vec{k}^2)^n+m^z]}{k^2_0+[(-\vec{k}^2)^n+m^z]^2},
\eea
and for $z=2n+1$,
\bea\label{zodd}
\frac{m^z}{g}=-2\int\frac{dk_0d^2\vec{k}}{(2\pi)^3}\frac{m^z}{k^2_0+(\vec{k}^2)^z+m^{2z}}.
\eea
Then, because of the integral $\int\frac{dk_0}{2\pi}\frac{1}{k^2_0+A^2}=\frac{1}{2A}$, we find that for the even $z$, we get the integral of a constant which vanishes within the dimensional regularization. Thus, for $z=2n$, we have $m^z=0$, so, there is no consistent vacuum shift, no mass term, and hence, no dynamical parity breaking in this case. At the same time, at $z=2n+1$ one has (here, the factor $\frac{1}{2}$ from the integral over $k_0$ is cancelled by the factor 2 from the trace of the $2\times 2$ unit matrix):
\bea
\frac{1}{g}=-\int\frac{d^2\vec{k}}{(2\pi)^2}\frac{1}{\sqrt{\vec{k}^{2z}+m^{2z}}},
\eea
which, for odd $z$, relates mass with the coupling $g$ through the equation
\bea
\label{oddgap}
m^{z-2}=-\frac{g}{4\pi^{3/2}}\Gamma\left(1+\frac{1}{z}\right)\Gamma\left(\frac{1}{2}-\frac{1}{z}\right).
\eea
It is clear that for any $z>1$ this equation is singularity-free. Actually, this expression implies that $g<0$, for any $z>1$.

Let us now verify that we indeed have a minimum of the effective potential. To do it, we proceed in the manner similar to \cite{Rosenstein}. The derivative of the effective potential, with the mass $m^z$ incorporated into the $\sigma$ field, is given by the expression
\bea
\frac{dV_{eff}}{d\sigma}=-\frac{\sigma}{g}-i\,\mathrm{tr} \int \frac{dk_0d^2\vec{k}}{(2\pi)^3} \frac{1}{\gamma^0k_0+(\gamma^ik_i)^z-\sigma}.
\eea
The similar calculations have been performed above with the only difference that $m^z$ instead of $\sigma$ was used. So, we can directly apply the result given in (\ref{oddgap}) for our case, and arrive at
\bea
\frac{\pa V_{eff}}{\pa\sigma}=-\frac{\sigma}{g}-\frac{\sigma^{2/z}}{4\pi^{3/2}}\Gamma\left(1+\frac{1}{z}\right)\Gamma\left(\frac{1}{2}-\frac{1}{z}\right).
\eea
This expression can be integrated, so that the result is
\bea
V_{eff}=-\frac{\sigma^2}{2g}-\frac{\sigma^{1+2/z}}{4\pi^{3/2}(1+2/z)}\Gamma\left(1+\frac{1}{z}\right)\Gamma\left(\frac{1}{2}-\frac{1}{z}\right).
\eea

In the above expression, $[g]=z-2$, $[\sigma]=z$, and, consequently, $[V_{eff}]=z+2$. Then, by rescaling $g$, $\sigma$, and consequently, $V_{eff}$, by the rules $g\to M^{z-1}g$, $\sigma\to M^{z-1}\sigma$, and $V_{eff}\to M^{z-1}V_{eff}$, in order to restore the usual dimensions of fields and couplings, with $M$ is a some arbitrary mass scale, we obtain 
\bea
V_{eff}=-\frac{\sigma^2}{2g}-\frac{M^{2(1-1/z)}\sigma^{1+2/z}}{4\pi^{3/2}(1+2/z)}\Gamma\left(1+\frac{1}{z}\right)\Gamma\left(\frac{1}{2}-\frac{1}{z}\right),
\eea
where now $[g]=-1$, $[\sigma]=1$, and $[V_{eff}]=3$. It is clear that for any odd $z\geq 3$ this effective potential will possess the minimum for a non-zero $\sigma$ (remind that $g<0$).

\subsection{Gap equation in the finite temperature case}

The above consideration can be generalized to the finite temperature case as well. In the finite temperature case the gap equation, in our theory for odd $z$,  is given by the Eq.~(\ref{zodd}). Then, we introduce the finite temperature through the Matsubara formalism.
Afterward, the Eq. (\ref{zodd}) takes the form
\begin{equation}\label{zodd2}
\frac{1}{g}=-\frac{2}{\beta}\sum_{l=-\infty}^{\infty}\intop\frac{d^2\vec{k}}{(2\pi)^{2}}\frac{1}{\omega_{l}^{2}+(\vec{k}^2)^z+m^{2z}},
\end{equation}
where $\omega_{l}=(2l+1)\frac{\pi}{\beta}$. Here $\beta$ is an inverse temperature, i.e., $\beta=1/T$, with its canonical mass dimension being equal to $-z$.

 Let us first evaluate the integral in (\ref{zodd2}), by using the solution
\bea
\intop\frac{d^{2}\vec{k}}{(2\pi)^{2}}\frac{1}{\omega_{l}^{2}+(\vec{k}^2)^z+m^{2z}} = \frac{1}{4\pi}\Gamma\left(1+\frac{1}{z}\right)\Gamma\left(1-\frac{1}{z}\right)\left(\omega_{l}^{2}+m^{2z}\right)^{\frac{1}{z}-1}.
\eea
Thus, we get the gap equation
\bea
\frac{1}{g}=-\frac{1}{2\pi\beta}\Gamma\left(1+\frac{1}{z}\right)\Gamma\left(1-\frac{1}{z}\right)\sum_{l=-\infty}^{\infty}\left(\omega_{l}^{2}+m^{2z}\right)^{\frac{1}{z}-1}.
\eea
Now, in order to carry out the above summation, let us use the expression \cite{Ford}
\begin{equation}\label{Ford}
\sum_{l=-\infty}^{\infty}\left[\left(l+b\right)^{2}+a^{2}\right]^{-\lambda}=\frac{\sqrt{\pi}\varGamma(\lambda-1/2)}{\varGamma(\lambda)(a^{2})^{\lambda-1/2}}+4\sin(\pi\lambda)f_{\lambda}(a,b),
\end{equation}
valid for $\lambda<1$, aside from the poles at $\lambda=1/2,-1/2,-3/2,\cdots$, with
\begin{equation}
f_{\lambda}(a,b)=\int_{|a|}^{\infty}\frac{dx}{(x^{2}-a^{2})^{\lambda}}Re\left(\frac{1}{e^{2\pi(x+ib)}-1}\right).\label{25}
\end{equation}
Note that, here, we have $a=\frac{\beta}{2\pi}m^{z}$, $b=\frac{1}{2}$, and $\lambda=1-\frac{1}{z}$. Thus, for our case (odd $z$) $\lambda$ always takes values $\lambda=0,\frac{2}{3},\frac{4}{5},\frac{6}{7},...,\frac{2n}{2n+1}$, i.e., between $0$ and $1$, so that we can readily use the relation (\ref{Ford}).
Hence, the integral (\ref{25}) converges and can be rewritten as
\bea
\label{soma}
f_{\lambda}\left(a,\frac{1}{2}\right)=\int_{|a|}^{\infty}dx\frac{\left(\tanh(\pi x)-1\right)}{2(x^{2}-a^{2})^{\lambda}}.
\eea

Finally, we have the following expression relating the coupling $g$ with the mass and the temperature:
\bea\label{get}
\frac{1}{g}&=&-
\left[\frac{m^{2-z}}{4\pi^{3/2}}\Gamma\left(1+\frac{1}{z}\right)\Gamma\left(\frac{1}{2}-\frac{1}{z}\right)+\frac{2\beta^{1-2/z}}{z(4\pi^2)^{1-1/z}}f_{1-1/z}\left(\frac{m^z\beta}{2\pi},\frac{1}{2}\right)
\right].
\eea
It is evident that the temperature independent terms of this expression replay the zero-temperature gap equation (\ref{oddgap}). Therefore, we find that the gap equation is modified in the finite temperature case. We note that, as the temperature is raised (or, equivalently, $\beta$ is decreased), the second term in the brackets in (\ref{get}), which is always a negative one, will approach the first term, so that they cancel each other at the critical temperature 
\bea\label{Tc}
T_c = \left[\frac{2^{-\frac{z+2}{z}} \pi ^{\frac{1}{2}-\frac{2}{z}} \Gamma \left(\frac{1}{2}-\frac{1}{z}\right) \Gamma \left(\frac{1}{z}\right)}{|f_{1-1/z}|}\right]^{\frac{z}{2-z}}m^z,
\eea
where we have considered $f_{1-1/z}=-|f_{1-1/z}|$. To find this critical temperature, we must approximate $f_{1-1/z}$ by its asymptotic value in the high temperature limit, i.e., we must calculate $f_{1-1/z}\left(a\to0,\frac12\right)$. For example, by taking into account $z=1$, we get $T_c=\frac{\sigma_0}{2\ln2}$, with $\sigma_0=-m$, which is the seminal result discussed in \cite{GNR}, where we have obtained $f_0=-\frac{\ln2}{2\pi}$, in the limite of $a\to0$.

In the following, for $z=3$, we have
\bea
T_c = \frac{16 \left(1-2^{2/3}\right)^3 \zeta \left(\frac{1}{3}\right)^3}{\sqrt{\pi } \Gamma \left(\frac{1}{6}\right)^3}m^3,
\eea
where $\zeta(\frac13)$ is the zeta function. To obtain this result, we have used the fact that, in the  high temperature limit,
\bea
f_{2/3}= -\frac{\left(1-2^{2/3}\right) \zeta \left(\frac{1}{3}\right) \Gamma \left(\frac{1}{3}\right)}{\sqrt[3]{2 \pi }}.
\eea
Thus, we see that there exists some critical temperature at which the parity breaking vanishes.

\subsection{Effective propagator and renormalizability}

To study further the dynamics, let us take into account propagator of the $\sigma$ field. It is characterized by the Feynman diagram given at Fig. 2.

\begin{figure}[htbp] 
\includegraphics[scale=0.8]{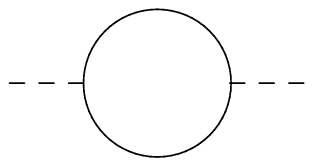}
\caption{\small{Contribution to the effective propagator of $\sigma$.}}
\label{figure2}
\end{figure}

The contribution of this graph, for even $z=2n$, looks like
\bea
\Gamma_2(p)=-\frac{N}{2}{\rm tr}\int\frac{dk_0d^{2}\vec{k}}{(2\pi)^3}\frac{(\gamma^0k_0-\omega(\vec{k}))(\gamma^0(k_0+p_0)-\omega(\vec{k}+\vec{p}))}{(k^2_0-\omega^2(\vec{k}))((p_0+k_0)^2-\omega^2(k+p))}
\sigma(p)\sigma(-p).
\eea
and for odd $z=2n+1$,
\bea
\Gamma_2(p)&=&-\frac{N}{2}{\rm tr}\int\frac{dk_0d^{2}\vec{k}}{(2\pi)^3}\frac{(\gamma^0k_0+\gamma^ik_i(-\vec{k}^2)^n+m^z)(\gamma^0(k_0+p_0)+\gamma^j(k_j+p_j)(-(\vec{k}+\vec{p})^2)^n+m^z)}
{(k^2_0-\Omega^2(\vec{k}))((k_0+p_0)^2-\Omega^2(k+p))}\nonumber\\
&&\times\sigma(p)\sigma(-p).
\eea
It is easy to calculate a trace in three dimensions since ${\rm tr}(\gamma^{\mu}\gamma^{\nu})=2\eta^{\mu\nu}$. So, for even $z=2n$, we have (here the mass is introduced by the formal reasons, since we note that there is no dynamical mass generation in this case; note nevertheless that either zero or non-zero mass does not affect the superficial degree of divergence of this theory since the mass never accompanies the UV leading orders of the propagators),
\bea
\Gamma_2(p)=-N\int\frac{dk_0d^{2}\vec{k}}{(2\pi)^3}\frac{k_0(k_0+p_0)+\omega(\vec{k})\omega(\vec{k}+\vec{p})}{(k^2_0-\omega^2(\vec{k}))((p_0+k_0)^2-\omega^2(\vec{k}+\vec{p}))}\sigma(p)\sigma(-p),
\eea
and, for odd $z=2n+1$,
\bea
\Gamma_2(p)&=&-N\int\frac{dk_0d^{2}\vec{k}}{(2\pi)^3}\frac{k_0(k_0+p_0)-\vec{k}\cdot(\vec{k}+\vec{p})(\vec{k}^2(\vec{k}+\vec{p})^2)^n+m^{2z}}
{(k^2_0-\Omega^2(\vec{k}))((k_0+p_0)^2-\Omega^2(k+p))}
\times\nonumber\\
&&\times\sigma(p)\sigma(-p).
\eea
It remains to calculate the integrals. 

The operator inverse to this effective action is just the propagator.
To integrate, we first do the Wick rotation, then use Feynman representation, with the subsequent change of variables $k_0\to k_0+p_0x$ is carried out for $k_0$ only. As a result, for even $z$, we get
\bea
\Gamma^{even}_2(p)&=&-iN\int\frac{dk_{0E}d^{2}\vec{k}}{(2\pi)^3}\int_0^1dx\frac{-k^2_{0E}+p^2_{0E}x(1-x)+\omega(\vec{k})\omega(\vec{k}+\vec{p})}{[k^2_{0E}+p^2_{0E}x(1-x)+(1-x)\omega^2(\vec{k})+x\omega^2(\vec{k}+\vec{p})]^2}\times\nonumber\\&\times&
\sigma(p)\sigma(-p),
\eea
and, for odd $z=2n+1$,
\bea
\Gamma^{odd}_2(p)&=&iN\int\frac{dk_{0E}d^{2}\vec{k}}{(2\pi)^3}\int_0^1dx[k^2_{0E}-p^2_{0E}x(1-x) 
+\vec{k}\cdot(\vec{k}+\vec{p})(\vec{k}^2(\vec{k}+\vec{p})^2)^n-m^{4n+2}]\nonumber\\
&&\times[k^2_{0E}+p^2_{0E}x(1-x)+(1-x)\Omega^2(\vec{k})+x\Omega^2(\vec{k}+\vec{p})]^{-2}
\sigma(p)\sigma(-p).
\eea
It remains to integrate over $k_{0E}$ and expand numerators and denominators in external momenta $(p_0,\vec{p})$, up to leading orders.

For even $z=2n$, after integration over $k_{0E}$ and re-introducing the Minkowski zero component through $p^2_{0E}\to -p^2_0$ we arrive at the following self-energy tensor $T(p)$:
\bea
T_{even}(p)=-\frac{iN}{4}\int\frac{d^{2}\vec{k}}{\left(2\pi\right)^{2}}\int_{0}^{1}dx\frac{x\omega^{2}(\vec{k}+\vec{p})+(1-x)\omega^{2}(\vec{k})+\omega(\vec{k})\omega(\vec{k}+\vec{p})+2x^{2}p_{0}^{2}}{[x\omega^{2}(\vec{k}+\vec{p})+(1-x)\omega^{2}(\vec{k})-x(1-x)p_{0}^{2}]^{\frac{3}{2}}},
\eea
which corresponds to $\Gamma_2$ through the relation $\Gamma_2=\sigma(-p)T(p)\sigma(p)$. 

The integral over $d^2\vec{k}$ in the expression above can be calculated only in an approximate manner. In order to consider the UV leading asymptotics of these propagators, we approximate the denominators by higher orders in $\vec{k}$ and $\vec{p}$. 
For this, let us first use
\bea
\label{approx}
\omega^{2}(\vec{k}) &\approx& (\vec{k}^2)^{2n}+m^{4n}, \nonumber\\
\omega(\vec{k})\omega(\vec{k}+\vec{p}) &\approx& (\vec{k}^2)^{2n}+\left(-1\right)^{n}m^{2n}(\vec{p}^2)^{n}+m^{4n}, \nonumber\\
\omega^{2}(\vec{k}+\vec{p}) &\approx& (\vec{k}^2)^{2n}+(\vec{p}^2)^{2n}+2(-1)^{n}(\vec{p}^2)^{n}m^{2n}+m^{4n}. 
\eea

With use of (\ref{approx}) we can write
\bea
T_{even}(p)\approx-\frac{iN}{4}\int_{0}^{1}dx\int\frac{d^{2}\vec{k}}{\left(2\pi\right)^{2}}\frac{2(\vec{k}^2)^{2n}+x(\vec{p}^2)^{2n}+(1+2x)(-1)^{n}m^{2n}(\vec{p}^2)^{n}+2x^{2}p_{0}^{2}+2m^{4n}}{[(\vec{k}^2)^{2n}+x(\vec{p}^2)^{2n}+2x(-1)^{n}(\vec{p}^2)^{n}m^{2n}+m^{4n}-x(1-x)p_{0}^{2}]^{\frac{3}{2}}}.
\eea
Now, after integrating over $\vec{k}$, we find
\bea
T_{even}(p)&\approx & -\frac{iN}{8\pi}\int_{0}^{1}dx\frac{1}{n\sqrt{\pi}}\{m^{4n}+[(\vec{p}^2)^{2n}+2(-1)^{n}(\vec{p}^2)^{n}m^{2n}-p_{0}^{2}(1-x)]x\}^{\frac{1}{2n}-\frac{3}{2}}\nonumber\\
&&\times\Gamma\left(\frac{1}{2n}+1\right)\left\{ \Gamma\left(\frac{n-1}{2n}\right)\{m^{4n}+[(\vec{p}^2)^{2n}+2(-1)^{n}(\vec{p}^2)^{n}m^{2n}-p_{0}^{2}(1-x)]x\}\right.\nonumber \\
 & & \left.+n\Gamma\left(\frac{3}{2}-\frac{1}{2n}\right)[2m^{4n}+(\vec{p}^2)^{2n}+2p_{0}^{2}x^{2}+(\vec{p}^2)^{n}m^{2n}(1+2x)]\right\} .
\eea
Considering the behavior of this expression in the limit $p_0\to\infty$, $\vec{p}\to\infty$ in order to obtain the UV asymptotics of the propagator, and approximating the integrals over $x$ as numbers of the order of 1, and we see that the behaviour of this self-energy tensor, up to numerical factors of order of 1 accompanying any monomials (these factors are denoted here by $b_1$, $b_2$, $b_3$), is
\bea
T_{even}(p)\simeq N\{m^{4n}+[b_1(\vec{p}^2)^{2n}-b_2p_0^{2}+2b_3(-1)^n(\vec{p}^2)^{n}m^{2n}+\cdots]\}^{\frac{1}{2n}-\frac{1}{2}}.
\eea
This expression is approximate but sufficient to find the superficial degree of divergence. The dots are for subleading orders in momenta.

For the odd $z$ we have the following self-energy tensor:
\bea
T_{odd}(p)&=&-\frac{iN}{4}\int\frac{d^{2}\vec{k}}{\left(2\pi\right)^{2}}\int_{0}^{1}dx
[x\Omega^{2}(\vec{k}+\vec{p})+(1-x)\Omega^{2}(\vec{k})-\vec{k}\cdot(\vec{k}+\vec{p})(\vec{k}^2)^{n}(\vec{k}^2+2\vec{k}\cdot\vec{p}+\vec{p}^2)^{n}\nonumber\\
&&+m^{4n+2}+2x^{2}p_{0}^{2}]\times[x\Omega^{2}(\vec{k}+\vec{p})+(1-x)\Omega^{2}(\vec{k})-x\left(1-x\right)p_{0}^{2}]^{-\frac{3}{2}},
\eea
where $\Omega^{2}\left(\vec{p}\right)=(\vec{p}^2)^{2n+1}+m^{4n+2}$. Again, we approximate the integral by taking into account only the leading orders in $\vec{k}$ and $\vec{p}$, which implies to consider
\bea
\Omega^{2}(\vec{k})&=&(\vec{k}^2)^{2n+1}+m^{4n+2}, \\
\Omega^{2}(\vec{k}+\vec{p})&\approx&(\vec{k}^2)^{2n+1}+(\vec{p}^2)^{2n+1}+m^{4n+2},\nonumber
\eea
and $\vec{k}\cdot(\vec{k}+\vec{p})(\vec{k}^2)^{n}(\vec{k}^2+2\vec{k}\cdot\vec{p}+\vec{p}^2)^{n}\approx(\vec{k}^2)^{2n+1}$. Therefore, the self-energy tensor is
\bea
T_{odd}(p)\approx-\frac{iN}{4}\int\frac{d^{2}\vec{k}}{\left(2\pi\right)^{2}}\int_{0}^{1}dx\frac{x(\vec{p}^2)^{2n+1}+2m^{4n+2}+2x^{2}p_{0}^{2}}{[x(\vec{p}^2)^{2n+1}+(\vec{k}^2)^{2n+1}+m^{4n+2}-x(1-x)p_{0}^{2}]^{\frac{3}{2}}}.
\eea
Integrating over $\vec{k}$, we arrive at
\bea
T_{odd}(p)\approx-\frac{iN}{8\pi^{\frac{3}{2}}}\int_{0}^{1}dx\Gamma\left(\frac{6n+1}{4n+2}\right)\Gamma\left(\frac{2n+2}{2n+1}\right)\frac{x(\vec{p}^2)^{2n+1}+2m^{4n+2}+2x^{2}p_{0}^{2}}{\left(m^{4n+2}+\left[(\vec{p}^2)^{2n+1}+p_{0}^{2}\left(x-1\right)\right]x\right)^{\frac{6n+1}{4n+2}}}.
\eea
Again, we consider the UV limit. We find
\bea
T_{odd}(p)\simeq N[(\vec{p}^2)^{2n+1}-a_1p_0^2+a_2m^{4n+2}+\cdots]^{\frac{1-2n}{2+4n}}.
\eea
Here and further, $a_1,a_2,b_1,b_2,\alpha,\beta$ are numerical factors, and dots are for terms with lower orders in momenta.

These results can be used to find effective propagators, and, finally, conclude about renormalizability of our theory. For even and odd $z$ respectively we have the propagators of $\sigma$ of the form $G=T^{-1}$ looking like 
\bea
G_{odd}(p)&=&T_{odd}^{-1}\propto\frac{1}{N}\frac{\alpha}{[(\vec{p}^2)^{2n+1}-a_1p_0^{2}+a_2m^{4n+2}]^{\frac{1-2n}{2+4n}}},\nonumber\\
G_{even}(p)&=&T_{even}^{-1}\propto\frac{1}{N}\frac{\beta}{[m^{4n}+b_1(\vec{p}^2)^{2n}-b_2p_0^2]^{\frac{1}{2n}-\frac{1}{2}}}.
\eea
Here we disregarded subleading orders in momenta.
We note that the effective dimension of $p_0$ is $2n+1$ for the odd case, and $2n$ for the even case, as it should be. It allows us to calculate the superficial degree of divergence $\omega$ of an arbitrary Feynman diagram.

For the odd $z=2n+1$, any propagator of $\psi$ contributes to $\omega$ with $-(2n+1)$, and any propagator of $\sigma$ -- with $2n-1$ (the non-negative contribution of propagators to the superficial degree of divergence is rather typical for the effective dynamics emerging due to quantum corrections, cf. \cite{sigma}). Any loop integration yields $2+z=2n+3$. Totally we have in this case
\bea
\omega=(3+2n)L+(2n-1)P_{\sigma}-(2n+1)P_{\psi},
\eea 
where $P_{\sigma}$ and $P_{\psi}$ are numbers of corresponding propagators. Using the topological identity $L+V-(P_{\sigma}+P_{\psi})=1$ together with relations between numbers of vertices and propagators $2P_{\psi}=2V-E_{\psi}$, $2P_{\sigma}=V-E_{\sigma}$, where $E_{\psi,\sigma}$ are numbers of corresponding external lines, we have
\bea
\omega=3+2n-E_{\psi}-(2n+1) E_{\sigma}.
\eea
Thus, the theory is renormalizable  in all orders of $\frac{1}{N}$ expansion, which matches the $z=1$  ($n=0$) result found in \cite{GNR}.

For the even $z=2n$, any propagator of $\psi$ contributes to $\omega$ with $-2n$, and any propagator of $\sigma$ -- with $2n-2$. In this case we have
\bea
\omega=(2+2n)L-2nP_{\psi}+(2n-2)P_{\sigma},
\eea
which yields
\bea
\omega=2+2n-E_{\psi}-2n E_{\sigma}.
\eea
Thus, for the even $z$ the theory is also renormalizable in all orders of $\frac{1}{N}$ expansion.

\section{Generation of the Chern-Simons term}

Now, to illustrate other possible impacts of HL-like spinor field, let us make the first step in studying of its interaction with a vector field, that is, 
let us couple the spinor field to a gauge field through the vertex
\bea
\label{int}
{\cal L}_{int}=\frac{e}{\sqrt{N}}\sum\limits_{j=1}^N\bar{\psi}_j\gamma_{\mu}\psi_j A^{\mu}.\label{14}
\eea
We note that this vertex is really a ``toy" coupling which we use only to discuss the possibility for a dynamical parity breaking similarly to the study performed in the relativistic case in \cite{GNR}, therefore, the Lagrangian given by the sum of (\ref{act}) and (\ref{int}) is really not gauge invariant, while to construct the complete model one should use the Lagrangian like $\bar{\psi}(i\gamma^0D_0-(i\gamma^iD_i)^z-m^z)\psi$, where $D_{0,i}=\pa_{0,i}-igA_{0,i}$ are the gauge covariant derivatives. That model was studied in the paper \cite{ourQED}, and we plan to study generation of the CS term in it in our next paper, while at this step our aim consists only in discussion of the parity breaking for which we use only the sum of (\ref{act}) and (\ref{int}).
  
Here, our aim is to find the polarization tensor given by
\begin{equation}
\Pi^{\mu\nu}(p)=e^{2}\int\frac{dk_{0}d^{2}\vec{k}}{(2\pi)^{3}}\mathrm{tr}\,\gamma^{\mu}S_{F}(k\text{+}p)\gamma^{\nu}S_{F}(k).\label{15}
\end{equation}

The corresponding Feynman diagram is given at Fig. 3.

\begin{figure}[htbp] 
\includegraphics[scale=0.8]{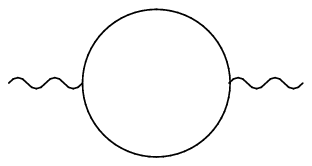}
\caption{\small{Contribution to the Chern-Simons term.}}
\label{figure3}
\end{figure}

We consider only the case of odd $z=2n+1$ since for $z=2n$ there is no mass generation and hence no CS term can arise.


For odd $z$ we use the propagator (\ref{3}), so that
(\ref{15}) yields
\bea
\Pi^{\mu\nu}(p)&=&-e^{2}\int\frac{dk_{0}d^{2}\vec{k}}{(2\pi)^{3}}\mathrm{tr}\,\gamma^{\mu}\frac{\gamma^{0}(k_{0}+p_{0})+\gamma^{i}(k_{i}+p_{i})(-\vec{k}^2-2\vec{k}\cdot\vec{p}-\vec{p}^2)^{n}+m^{z}}{\left(k_{0}+p_{0}\right)^{2}-(\vec{k}^2+2\vec{k}\cdot\vec{p}+\vec{p}^2)^{z}-m^{2z}}\nonumber\\
&&\times\gamma^{\nu}\frac{\gamma^{0}k_{0}+\gamma^{i}k_{i}(-\vec{k}^{2})^{n}+m^{z}}{k_{0}^{2}-(\vec{k}^2)^{z}-m^{2z}},
\eea
Here we look for the CS term. To find it, we concentrate on terms linear in the external momentum $p$ and involving three Dirac matrices, to get the Levi-Civita tensor. The relevant terms are
\bea
\Pi^{\mu\nu}_{CS}&=&-e^{2}\int\frac{dk_{0}d^{2}\vec{k}}{(2\pi)^{3}}\mathrm{tr}\frac{\gamma^{\mu}\gamma^{0}p_{0}\gamma^{\nu}m^{z}+\gamma^{\mu}\gamma^{i}p_{i}\gamma^{\nu}(-\vec{k}^{2})^{n}m^{z}}{[(k_{0}+p_{0})^{2}-(\vec{k}^2+2\vec{k}\cdot\vec{p}+\vec{p}^2)^{z}-m^{2z}][k_{0}^{2}-(\vec{k}^2)^{z}-m^{2z}]}.
\eea
Now, by calculating the traces, we arrive at
\bea
\Pi^{\mu\nu}_{CS}&=&2im^{z}e^{2}\int\frac{dk_{0}d^{2}\vec{k}}{(2\pi)^{3}}\frac{\epsilon^{\mu0\nu}p_{0}+\epsilon^{\mu i\nu}p_{i}(-\vec{k}^{2})^{n}}{[(k_{0}+p_{0})^{2}-(\vec{k}^2+2\vec{k}\cdot\vec{p}+\vec{p}^2)^{z}-m^{2z}][k_{0}^{2}-(\vec{k}^2)^{z}-m^{2z}]}.
\eea
The low-energy leading terms are
\bea
\Pi^{\mu\nu}_{CS}&=&-2m^{z}e^{2}\int\frac{dk_{0E}d^{2}\vec{k}}{(2\pi)^{3}}\frac{\epsilon^{\mu0\nu}p_{0}+\epsilon^{\mu i\nu}p_{i}(-\vec{k}^{2})^{n}}{[k^2_{0E}+(\vec{k}^2)^{z}+m^{2z}]^2},
\eea
where we have performed the Wick rotation. After the integration over $k_{0E}$, we obtain
\bea
\Pi^{\mu\nu}_{CS}&=&-\frac{m^{z}e^{2}}{2}\int\frac{d^{2}\vec{k}}{(2\pi)^{2}}\frac{\epsilon^{\mu0\nu}p_{0}+\epsilon^{\mu i\nu}p_{i}(-\vec{k}^{2})^{n}}{[(\vec{k}^2)^{z}+m^{2z}]^{\frac32}}.
\eea
Finally, by calculating the integral over $\vec{k}$, we get
\bea
\Pi^{\mu\nu}_{CS}&=&-\frac{e^{2}}{4\pi^{3/2}}\mathrm{sgn}(m)(c_{1}\varepsilon^{\mu0\nu}p_{0}+c_{2}\varepsilon^{\mu i\nu}p_{i}),\label{17}
\eea
where
\bea
c_{1}&=&m^{-4n}\,\Gamma\left(\frac{2n+2}{2n+1}\right)\Gamma\left(\frac{6n+1}{4n+2}\right),\\
c_{2}&=&m^{-2n}\frac{(-1)^n}{n+1}\,\Gamma\left(\frac{3n+2}{2n+1}\right)\Gamma\left(\frac{4n+1}{4n+2}\right).\nonumber
\eea

The Eq. (\ref{17}) is a self-energy tensor describing generating the CS term in the Gross-Neveu model at odd $z$. 
It is easy to see that for  $z\neq1$
one has $c_{1}\neq c_{2}$ which confirms the space-time anisotropy.
To verify the consistency of our result, we note that at $z=1$, that is, $n=0$, we get
\bea
\Pi^{\mu\nu}_{CS}&=&-\frac{e^{2}}{4\pi}\mathrm{sgn}(m)\epsilon^{\mu\rho\nu}p_{\rho}.\label{18}
\eea
The Eq. (\ref{18}) is just the same obtained in \cite{GNR}, so we see that at $z=1$ the Lorentz invariance is recovered as it should be.

\section{Summary}

We considered different issues related with the Horava-Lifshitz-like $(2+1)$-dimensional Gross-Neveu model. We showed that within the framework of the $\frac{1}{N}$ expansion, it is renormalizable for any $z$. Also, we demonstrated (although the vector field was introduced from the beginning as the external one, just as in \cite{Dhar}, and not generated through some Thirring-like interaction) that the anisotropic Chern-Simons term can be generated in it (however, we note that, for the simplified model we considered here, it is not gauge invariant). A consequence of generating the Chern-Simons term consists in the fact that we succeeded in this way to break the parity dynamically. Besides, we succeeded to generalize some of these results to the finite temperature case, namely, we considered the gap equation in the finite temperature case and found that there is some critical temperature at which the parity breaking disappears (in principle, one could also apply this result to detailed study of possible phase transitions in this theory extending thus the studies carried out in \cite{Dhar}).

We showed that our emergent dynamics of the Lagrange multiplier $\sigma$ is renormalizable for any $z$ (unlike \cite{prsmg}, to prove it we do not use any special identities). Therefore our HL-like Gross-Neveu model is perturbatively consistent. 

The possible continuation of our study could consist in studying of the emergent dynamics of vector fields in more details, that is, generation of the vector field through Thirring-like interaction in a manner similar to that one used here to introduce the Lagrange multiplier $\sigma$, and generating of the CS term starting with the full-fledged, gauge invariant Lagrangian $\bar{\psi}(i\gamma^0D_0-(i\gamma^iD_i)^z-m^z)\psi$ \cite{ourQED}. This is a problem we plan to consider in the next paper.

{\bf Acknowledgements.}   R. Mart\'{i}nez would like to thank to Francisco Pe\~{n}a for the lectures on QFT at finite temperature. This work was partially supported by Conselho
Nacional de Desenvolvimento Cient\'{\i}fico e Tecnol\'{o}gico (CNPq). The work by A. Yu. P. has been supported by the CNPq project 303783-2015/0.

\end{document}